\documentclass[12pt]{article}

\usepackage{epsfig,graphics,amssymb,here}

\newcommand{\be} {\begin{equation}}
\newcommand{\ee} {\end{equation}}
\newcommand{\bma} {\begin{math}}
\newcommand{\ema} {\end{math}}
\newcommand{\beqa} {\begin{eqnarray}}
\newcommand{\eeqa} {\end{eqnarray}}
\newcommand{\ga} {\gamma}

\newcommand{\nn} {\nonumber}
\def\vec#1{\mathchoice{\mbox{\boldmath$\mathrm\displaystyle#1$}}
{\mbox{\boldmath$\mathrm\textstyle#1$}}
{\mbox{\boldmath$\mathrm\scriptstyle#1$}}
{\mbox{\boldmath$\mathrm\scriptscriptstyle#1$}}}
\newcommand{\bm}[1]{\mbox{\boldmath$#1$}}  
\renewcommand{\vec}{\bm}
\newcommand{\simgt}{\hbox{ \raise3pt\hbox to 0pt{$>$}
    \raise-3pt\hbox{$\sim$} }}
\newcommand{\simsm}{\hbox{ \raise3pt\hbox to 0pt{$<$}
    \raise-3pt\hbox{$\sim$} }}
\begin{document}

\begin{flushright}
HD--THEP--99--3\\
TAUP--2548--99
\end{flushright}

\vspace{0.2cm}

\begin{center}

{\LARGE{ Odderon and photon exchange in 
electroproduction of pseudoscalar mesons}\footnote{Supported 
by German 
Bundesministerium f\"ur Bildung und Forschung (BMBF),\\
\hphantom{dfd} Contract Nr. 05 7HD 91 P(0),
by Deutsche Forschungsgemeinschaft
under grant\\ 
\hphantom{dfd} no. GKR 216/1-98, by PPARC,
by DAAD 
and MINERVA-Stiftung.\\[0.06cm]
\hphantom{dfde}$email\; adresses$: 
  e.berger@thphys.uni-heidelberg.de,
  ad@a3.ph.man.ac.uk,\\
  \hphantom{dfde}h.g.dosch@thphys.uni-heidelberg.de,
  kilian@particle.physik.uni-karlsruhe.de,\\
  \hphantom{dfde}o.nachtmann@thphys.uni-heidelberg.de,
  rueter@post.tau.ac.il.} }

\vspace{0.1cm}

\end{center}

\begin{center}

{\large 
E. R. Berger${}^{\rm a}$, 
A. Donnachie${}^{\rm b}$
H. G. Dosch${}^{\rm a}$,
W. Kilian${}^{\rm c}$,
O. Nachtmann${}^{\rm a}$,
M. Rueter${}^{\rm d}$ }
\end{center}

\vspace{0.1cm}

$\hphantom{ss\;ss}{}^{{\rm a}}$ : 
$\, \;$Institut  f\"ur  Theoretische Physik
                der Universit\"at Heidelberg,\\ 
                \hphantom{ : ${}^{1,3,5}$sss}
                Philosophenweg 16, 
                D-69120 Heidelberg
\newline
\hphantom{ssssss}${}^{\rm b}$ : 
         $ \,$TH Division, CERN;\\ 
         \hphantom{aa${}^{1}\;\,$ssssss}Permanent address:
         Department of Physics and Astronomy,\\
\hphantom{aa${}^{1}$ssssss}
         University of Manchester, Manchester M13 9PL, UK 
\newline
\hphantom{ssssss}${}^{\rm c}$ : $ \;$Institut f\"ur Theoretische 
         Teilchenphysik,\\
         \hphantom{aa${}^{1}$ssssss}
         Universit\"at Karlsruhe,
         D-76128 Karlsruhe
\newline
\hphantom{ssssss}${}^{\rm d}$ : $ \;$School of Physics and Astronomy,
         Department of High Energy Physics,\\
         \hphantom{aa${}^{1}$ssssss}
         Tel Aviv University,
         69978 Tel Aviv, Israel

\vspace{0.5cm}

\thispagestyle{empty}

\begin{abstract}

We investigate the reaction 
${\rm e}\;{\rm p} \rightarrow {\rm e}\;{\rm PS}\;{\rm X}$
where PS denotes a pseudoscalar meson 
$\pi^0$, $\eta$, $\eta'$, or $\eta_c$
and X either a proton or resonance or continuum states
into which the proton can go by diffractive
excitation. At high energies photon and odderon exchange 
contribute to the reaction. The photon exchange contribution 
is evaluated exactly using data for the total virtual
photon-proton absorption cross section. The odderon exchange
contribution is calculated in nonperturbative QCD,  using
functional integral techniques and the model of the stochastic 
vacuum. For the proton we assume a quark-diquark 
structure as suggested by the
small odderon amplitude in ${\rm pp}$ and $ {\rm p} \bar{{\rm p}}$
forward scattering. We show that odderon exchange leads to 
a much larger inelastic than elastic PS production cross section.
Observation of our reaction at HERA would establish the
soft odderon as an exchange object on an equal footing with the soft 
pomeron and would give us valuable insight into both the 
nucleon structure and the mechanism of high energy diffractive 
scattering.

\end{abstract}

\newpage

\section{Introduction}

The phenomenological (non-perturbative) pomeron has long been established
as an effective Regge pole whose exchange governs high-energy diffractive
hadron-hadron scattering \cite{dola}. 
The pomeron carries vacuum quantum numbers 
$C = P = +1$, and there is no {\it a priori} reason why a $C = P = -1$
partner, the phenomenological odderon \cite{nico1},  
should not exist. If present, 
the exchange of an odderon Regge pole would produce a difference between 
${\rm p p}$ and ${\rm p}\bar {\rm p}$ 
scattering at high energies and at small momentum 
transfer. A particularly sensitive test is provided by measurements of 
the forward real part of the ${\rm p p}$ and ${\rm p}\bar {\rm p}$
scattering amplitudes, 
and they are consistent with the absence of odderon exchange \cite{auge}.
There are two possible explanations for 
the apparant absence of the odderon. One is that the non-perturbative
odderon really does not exist. This seems implausible as QCD-based models 
of the phenomenological pomeron can easily be extended to describe a 
phenomenological odderon. Further a perturbative ``odderon'', namely 
three-gluon exchange\footnote{A lot of theoretical work has been 
devoted to calculating perturbative QCD (pQCD) corrections to this type
of odderon \cite{pod1,pod2,pod3,pod4}. The result seems to be that such
corrections have a small effect, for instance changing the (effective)
intercept of the odderon trajectory $\alpha_{\mathbb{O}}(0)$ 
by less than 10$\%$ 
\cite{pod2,pod4,pod5} from the value of 1, which is the result for the 
lowest order three gluon exchange.} 
is believed to dominate large-angle ${\rm p p}$ and ${\rm p}\bar {\rm p}$ 
scattering. One of the most compelling arguments for a $C = -1$ exchange in 
high energy scattering is provided by the ${\rm p p}$ and 
${\rm p}\bar {\rm p}$ differential 
cross sections at the ISR where a deep dip in the former process is 
transformed into a shoulder in the latter. The second possibility is that 
the phenomenological odderon does exist, but that its coupling to the nucleon 
in elastic scattering at small $t$ is extremely small. That is the view we 
take here.

One successful approach to high-energy diffractive scattering is based on 
functional integral techniques \cite{lana,na91}
and the use of the model of the stochastic 
vacuum (MSV) \cite{msv}
to evaluate the correlation functions of the Wegner-Wilson 
loops which occur in the formalism when applied to 
hadron-hadron scattering \cite{dfkpre,dfk}. 
This gives a remarkably good description 
of many different processes involving the exchange of vacuum quantum
numbers \cite{dfk,all,dogupi,doku,naber}.
This model can easily be extended to the exchange of a $C = P = -1$ object,
and it has been shown that the clustering of two quarks to form a diquark in
a nucleon leads to a drastic reduction in the odderon-$N$-$N$
coupling \cite{doruod}.
The general result is
rather model independent. It relies on the fact that the quark-diquark
density in a nucleon is nearly symmetric under a parity transformation 
(if the diquark is sufficiently small) whereas the odderon coupling changes
sign. Therefore there is a cancellation when the nucleon wave function is
integrated over all angles. Specifically, in the MSV it has been shown that 
the formation of a diquark with a radius $r_D < 0.3$ fm yields sufficient
suppression of the odderon coupling to the proton in order to be in agreement
with the upper limit allowed by the measurements of the forward real part of
the ${\rm p p}$ and ${\rm p}\bar {\rm p}$ scattering amplitudes.

It should be stressed that this argument is only valid if the odderon is
treated as a simple Regge pole near $J = 1$. If a more complicated situation
is permitted, as in \cite{nico2}, then the phenomenology is very different
and the suppression of odderon exchange in ${\rm p p}$ and 
${\rm p}\bar {\rm p}$ scattering 
at high energies is much less marked. However the data do not {\it require}
the additional complexity of double and triple poles, so until demonstrated
explicitly otherwise we prefer to stay with the simplest model which agrees
with data and which allows essentially parameter-free predictions. 

It is clearly advantageous if one can find high-energy reactions which permit
odderon exchange but exclude pomeron exchange. Exclusive neutral pseudoscalar 
meson production in ${\rm e p}$ 
scattering at high energies is one such, providing a
direct probe for odderon exchange \cite{schmna,engel}. 
Using a simple ansatz for the odderon, this
cross section has been calculated \cite{kina}. 
Because of the suppression of the 
odderon-p-p coupling, the cross section due to photon-photon fusion is
comparable to that expected at best from odderon exchange. Nonetheless there 
are kinematical distributions which could provide promising signals for the 
odderon.

The suppression of the odderon coupling due to diquark formation does not hold
if the nucleon is transformed diffractively into an excited negative parity 
state. In this case, even for a pointlike diquark the odderon couples to
the nucleon without any restriction \cite{donaru}
giving promise of a significantly higher
cross section for odderon exchange in, for example, 
${\rm e p} \rightarrow {\rm e} \pi^0 {\rm X}$ 
(where X stands for diffractively excited proton states) 
compared to ${\rm e p} \rightarrow {\rm e} \pi^0 {\rm p}$. 
In fact this is much more
representative of the real experimental situation, as it is often not 
possible to say whether the proton has recoiled quasi-elastically or has 
been transformed into an excited state within some experimentally defined 
mass range. At HERA the upper limit for the recoil mass $M_X$ is typically
$\sim 2$ GeV. Accordingly we calculate the contribution from  odderon 
exchange to pseudoscalar meson
production 
with nucleon fragmentation in ep interactions at high energies , Fig.1(a). 
To complement this we also calculate the contribution to the same process 
from photon exchange, Fig.1(b). This latter calculation is exact, limited 
only by the errors on the ep total cross section for the $(Q^2,M_X)$ 
range of relevance. When combined with the previous calculation 
\cite{kina} of photon 
exchange for the quasi-elastic process this provides an absolute
prediction for
the cross section from the electromagnetic
process alone. Any measurement deviating significantly from this would 
be strong evidence for the odderon. 
%
%
%
\begin{figure}[htb]
  \unitlength1.0cm
  \begin{center}
    \begin{picture}(15.,8.8)

      \put(-0.6,1.0){
        \epsfysize=5.0cm
        \epsffile{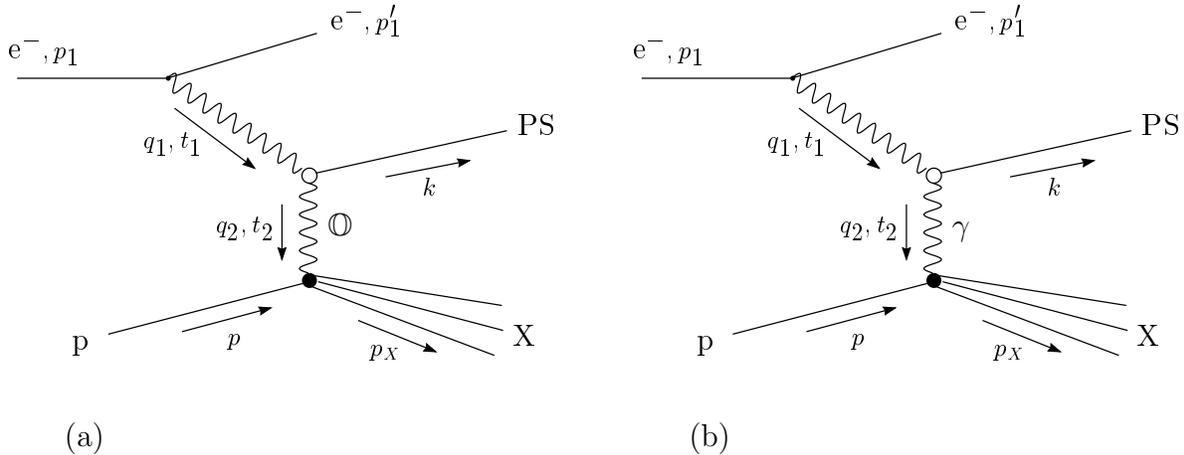}}

      \put(0.4,0){(a)}

      \put(7.7,1.0){
        \epsfysize=5.0cm
        \epsffile{pseudofeyn.eps}}

      \put(8.7,0){(b)}

      \put(3.9,2.8){$\mathbb{O}$}

      \put(12.2,2.8){$\gamma$}

    \end{picture}
  \end{center}
  \vspace*{-0.0cm}
  \caption{Feynman diagrams for pseudoscalar meson production in $ep$
scattering at high energies with odderon (a) and photon (b) exchange.}
\end{figure}
%
%
%
Thus we are considering electroproduction of a pseudoscalar meson
PS = $\pi^0,\eta,\eta',\eta_c$ with nucleon break-up: 
\beqa
  {\rm e}^{\pm}(p_1) + {\rm p}(p) 
  \rightarrow {\rm e}^{\pm}(p_1') + {\rm PS}(k) + {\rm X}(p_X).
\label{reaction}
\eeqa
%
%
We define $q_1 = p_1 - p_1'$, $q_2 = q_1-k = p_X - p$, 
$s=(p_1+p)^2$, $W=s_2=(q_1+p)^2$,
$t_1 =q_1^2=-Q^2$ 
and $t_2 = q_2^2$. Here we treat the very small $Q^2$ range. 
In the H1 experiment at HERA the kinematical cuts for this
so called photoproduction region \cite{hera} are
\beqa
  && y_{{\rm min}}=0.3 \le y \le 0.7=y_{{\rm max}},
  \nonumber\\
  && 0 < Q^2 < 0.01 \, {\rm GeV^2},
  \label{cuts}
\eeqa
%
%
where, in the proton rest frame $y = (p q_1)/(p p_1)$ is the fractional
energy loss of the incoming lepton.
Due to the cuts (\ref{cuts}) the photons emitted by the ${\rm e}^{\pm}$
are always nearly on shell, and the equivalent photon approximation
(EPA) \cite{budnev} is applicable.
The total electroproduction cross section for producing a PS
in terms of the EPA is given by
\begin{eqnarray}
  &&\sigma = \int_{y_{min}}^{y_{max}} \, \frac{dy}{y} \, n(y) \;
  \sigma_{\gamma {\rm p}}
  (s_2)
  \; \; \; \; \; {\rm with} \; \; \; \; \;  s_2=y s + (1-y) m_{{\rm p}}^2,
  \nonumber\\[0.2cm]
  &&n(y) = \frac{\alpha}{\pi} \, 
  \bigg\{ 
  \Big(1-y+\frac{y^2}{2}\Big) \, {\rm ln} \Big(
  \frac{|t|_u}{|t|_l}\Big) -
  \frac{m_e^2 y^2}{|t|_l} \, 
  \Big( 1 - \frac{|t|_l}{|t|_u} \Big) \ -
  \nonumber\\
  && \hphantom{N(y) = \frac{\alpha}{\pi} \, b }
  \Big( 1-\frac{y}{2} \Big)^2 \, 
  {\rm ln} \Big( \frac{ |t|_u/E^2 + y^2} { |t|_l/E^2 + y^2} \Big) \bigg\},
  \label{epa}
\end{eqnarray}
%
%
where $m_{{\rm p}}$ 
is the nucleon mass and $\sigma_{\gamma {\rm p}}$ is the total photoproduction
cross section for the reaction. $n(y)$ is the equivalent photon number 
\cite{budnev} for a given 
energy fraction  of the incoming ${\rm e}^{\pm}$ transfered
to the $\gamma {\rm p}$ subsystem. In the proton rest frame 
$E=p p_1/m_{{\rm p}}$
is the energy of the incoming $e^{\pm}$.
The upper limit $|t|_u$ is given by experiment (\ref{cuts}),  
$|t|_u$=0.01 ${\rm GeV}^2$, and $|t|_l$=$m_e^2 y^2/(1-y)$.

\section{The odderon contribution}

For convenience we discuss the odderon exchange contribution to the
reaction $\gamma {\rm p} \rightarrow \pi^0 {\rm X}$. 
The conversion to an incident
electron via the EPA is in principle straightforward, as is the replacement 
of the $\pi^0$ by one of the other pseudoscalars.

There are two apparently extreme approaches to this calculation. One
is to assume that the system X is dominated by a small number of
resonances, which for the odderon should be a good approximation.  The
second is to ignore any possible structure in the final system X and
represent it simply by a free quark-diquark pair. Both calculations are 
performed and give similar results.

We discuss first resonance production.
The break-up of the nucleon by odderon exchange leads naturally to negative 
parity final states. The lowest lying nucleonic isoscalar states with 
negative parity are the $N(1520)$: $J^{P}={\frac{3}{2}}^- $ 
and the $N(1535)$: 
$J^{P}={\frac{1}{2}}^-$. 
These are both compatible with the diquark picture. 
For dynamical reasons a scalar diquark is favoured over a vector one
\cite{shure}.
The P-wave excitation of a quark and a scalar diquark 
gives degenerate nucleon resonances with quantum numbers $J^P = 
{\frac{3}{2}}^-, {\frac{1}{2}}^-$ which can readily be identified with 
these two observed states. 

It is easy to see that considering unpolarised cross sections, 
summed over both resonances the quark spin degree of freedom becomes
irrelevant and the calculation reduces to one where a spinless state is 
exited to a 2P resonance.

The scattering amplitude $T(s_2,t_2)$, calculated in the
$\gamma {\rm p}$ c.m. system, can be expressed through a profile 
function $J(\vec b)$:

\be  \label{fund}
T(s_2,t_2) = 2is_2 \int \,
d^2b\,e^{i\vec{q_2}_T \vec{b}}\,J(\vec b).
\label{msvampl}
\ee
%
%

The profile function is expressed in the model \cite{dfk}
as an overlap of a
dipole-dipole scattering amplitude $\tilde J(\vec b, \vec r_1, z_1, \vec
r_2, z_2)$ where $\vec b$ is the impact parameter of two
lightlike dipole trajectories with (transverse) size $\vec r_1$ and
$\vec r_2$ respectively. The quantities $z_1$, $z_2$ are the
longitudinal momentum fractions of the quarks in the dipoles.

It is of course necessary to take a profile function corresponding 
to the exchange of a $C$=$P$=$-$1 object \cite{doruod}. Note that the 
model contains  only the kinematical $s_2$-dependence shown in  
(\ref{fund}) which leads to an energy independent cross section. The
parameters of the model\footnote{The parameters of the pion-photon
overlap can be found in \cite{donaru}. The MSV parameters are: $a$=0.31 fm,
$\langle g^2 \, FF \rangle$=3.0 ${\rm GeV}^4$, $S_{\rm p}$=0.85 fm as
determined in \cite{mdoc}.}
are the same as those used in \cite{donaru} which were 
fixed at an energy of  $\sqrt{s_2}$=$W$=20 GeV. We return to the question of 
energy dependence at the end of this section.

Hadronic amplitudes for production of hadronic resonance
states are obtained by smearing the dipole extensions
$\vec r_i$ with the respective transition densities: 
\beqa 
 J(\vec b)_{\lambda,\lambda_{\gamma}} &=& 
 \int \frac{d^2 r_1}{4\pi} dz_1
 \int \frac{d^2 r_2}{4\pi} 
 \nn\\
 && \sum_{f,h_1, h_2} 
 \Psi^{*\, \pi^0}_{f h_1 h_2}(\vec{r}_1,z_1)
 \Psi^{\ga}_{\lambda_{\gamma},\,f h_1 h_2} (\vec{r}_1,z_1)
 \Psi^{*\, {\rm 2P}}_{\lambda} (\vec{r}_2)
 \Psi^{{\rm p}}(\vec{r}_2)
 \tilde{J}(\vec b, \vec{r}_1,z_1, \vec{r}_2).
\label{profile}
\eeqa
%
%
Here $\lambda,\lambda_{\gamma}$ stands for the helicities of the
photon and of the the orbital helicity of the resonance, respectively. 
In agreement with the other applications of the model we use for the
quark-diquark wave function of the proton a Gaussian with an extension
parameter $S_{\rm p}$ adjusted to $pp$ scattering and neglect the
dependence on the longitudinal momentum fraction $z_2$ in the purely
hadronic overlap. Thus we take for the proton wave function
\be 
\label{proton}
\Psi^{\rm p}(r_2)=
\frac{\sqrt{2}e^{-{r_2^2}/{4S_{\rm p}^2}} }{S_{\rm p}}.
\label{pwave}
\ee
%
%
As the
orbital wave function for the low lying degenerate excited states we choose 
an ansatz analogous to the proton-wave function but in a P-state:
\beqa
&&\Psi^{{\rm 2P}}_{\lambda} (\vec{r}_2)=
 \Psi^{{\rm 2P}} (r_2) \; e^{i \lambda \theta_2},
 \nonumber\\
&& \Psi^{{\rm 2P}} (r_2) =  
 \frac{r_2 \; e^{-{r_2^2}/{4 S_{\rm p}^2}} }{ S_{\rm p}^2 } 
 \label{2pwave}
\eeqa
%
%
with the same extension parameter $S_{\rm p}$ as for the
proton. 
An analogous strategy has been applied successfully for the excited
$\rho$ states \cite{doku}.
In (\ref{msvampl}) we use $\vec{q_2}_T$ as x-axis
for the transverse vectors  $\vec{r}_1$, $\vec{r}_2$ and $\vec{b}$, so
$\theta_2$ is the angle of $\vec{r}_2$ in planar coordinates.  
The orbital helicity $\lambda$ of the 2P state can 
take the values $0,\pm 1$, but the orbital helicity 
$\lambda=0$ cannot be excited in our model.

The photon-pion overlap is taken from \cite{donaru}:
\beqa
&& \sum_{f,h_1, h_2} 
 \Psi^{*\, \pi^0}_{f h_1 h_2} (\vec{r}_1,z_1)
 \Psi^{\ga}_{\lambda_{\gamma},f h_1 h_2} (\vec{r}_1,z_1) = 
 \nn\\
&&i\frac{e}{\sqrt{2}} 
 f_\pi e^{-{\omega^2 r_1^2}/{2}} e^{i\lambda_{\gamma} \theta_2}
 z_1(1-z_1) f(z_1) \left(
 m_q \; {\rm K}_1 (m_q r_1) + 
 r_1 \omega^2{\rm K}_0 (m_q r_1) \right)
 \label{piwave}
\eeqa
%
%
Here $m_q$ is the quark mass, $m_q=0.22$ GeV.
For a justification of such a simple ansatz see \cite{dogupi}.
It should be noted that the
$\gamma\gamma$ decay width of the $\pi^0$ comes out correctly with this 
overlap. The function $f(z_1)$ is given by \cite{bsw}:
\be
f(z_1)= {\cal N} \sqrt{z_1(1-z_1)} 
\exp\left(
-\frac{M_\pi^2(z_1-1/2)^2}{2\omega^2}\right)
\ee
%
%
and $\omega$ and $\cal{N}$ are fixed by normalization.

\begin{figure}[htb]
  \unitlength1.0cm
  \begin{center}
    \begin{picture}(15.,9.8)
      \put(3.0,1.0){
        \epsfysize=9.0cm

        \epsffile{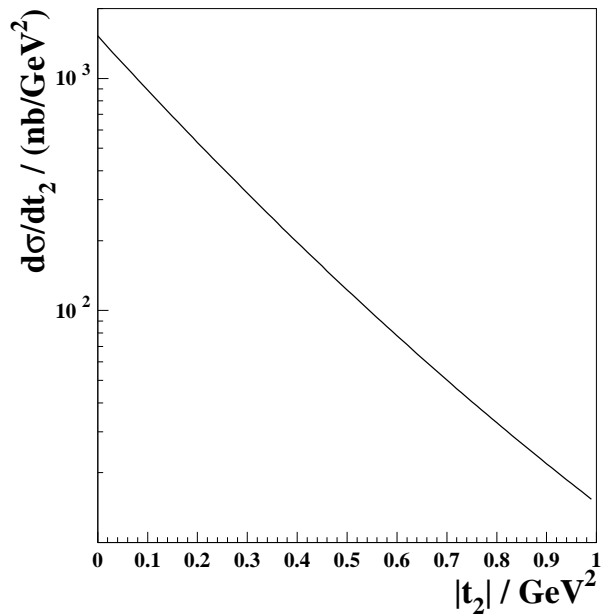}}
    \end{picture}
  \end{center}
  \vspace*{-1.5cm}
  \caption{The differential cross section 
     $d\sigma_{\gamma {\rm p}}^{\mathbb{O}}  / dt_2$ 
  of the process
  $\gamma {\rm p} \to \pi^0 \{{\rm 2P}\}$ as a function of $t_2$. 
(A fit to this curve is: 
$d\sigma_{\gamma {\rm p}}^{\mathbb{O}}  / dt_2$
=$a \, \exp( -b t_2 -c t_2^2)$ where 
$a = 1523 \; {\rm nb}, \;b = 5.44 {\rm GeV}^{-2},  
\; c = - 0.80 {\rm GeV}^{-4}$)}
  \label{msvbild}
\end{figure}

To take advantage of global azimuthal invariance
it is useful to choose as new integration variables
the relative angles between $\vec{b}$ and 
$\vec{r}_{1(2)}$
$\theta_{1(2)}^{\prime} = \theta_{1(2)}-\theta_b$. Then 
with this choice of coordinates $\tilde{J}$ 
in (\ref{profile}) becomes independent
of $\theta_b$
and inserting  (\ref{pwave}), (\ref{2pwave}) and (\ref{piwave}) into the
scattering amplitude (\ref{msvampl}) we get
\beqa
T_{\lambda , \lambda_{\gamma}}&=&2is_2 \int b\, db\,d\theta_b \,
 e^{i\sqrt{-t_2} b \cos (\theta_b)} 
 \int \frac{d^2 r_1}{4\pi} d z_1 
 \int \frac{d^2 r_2}{4\pi}
 \nn\\
&&\times 
 \Big[\sum_{f,h_1, h_2}\Psi^{*\, \pi^0}_{f h_1 h_2}(r_1,z_1)
 \Psi^{\ga}_{\lambda_{\gamma}, f h_1 h_2}(r_1,z_1)\Big]_{\theta_1=0}
 \Psi^{*\, 2P}(r_2)
 \Psi^{{\rm P}}(r_2)
 \nn\\
&&\times 
 e^{i \lambda_{\gamma} (\theta_1^{\prime}+\theta_b)} 
 e^{i \lambda (\theta_2^{\prime}+\theta_b)}
 \tilde{J}(\vec{b}, \vec{r}_1,z_1, \vec{r}_2),
\eeqa
%
%
where the wavefunctions now only depend on $r_{1(2)}$. 
Using the relation
\be
\int_0^{2\pi}d\theta_b \,e^{i\sqrt{-t_2} 
b \cos (\theta_b)} e^{in\theta_b}=(i)^n 2\pi 
J_n(\sqrt{-t_2}b)\nn
\ee
%
%
we can now perform the $\theta_b$ integration and 
obtain finally for the scattering amplitude
\beqa
 &&T_{\lambda,\lambda_{\gamma}} = 2 i s_2 \int b\, db\,
 \int \frac{d^2 r_1}{4\pi} dz_1
 \int \frac{d^2 r_2}{4\pi}
 \nn\\
&&\hphantom{T_{\lambda_{\gamma},\lambda}}
 \times 
 \Big[ \sum_{f,h_1, h_2}
 \Psi^{*\, \pi^0}_{f h_1 h_2}(r_1,z_1)
 \Psi^{\ga}_{\lambda_{\gamma},f h_1 h_2}(r_1,z_1) \Big]_{\theta_1=0}
 \Psi^{*\, 2{\rm P}}(r_2)\Psi^{{\rm p}}(r_2)
 \nn\\
&&\hphantom{T_{\lambda_{\gamma},\lambda_{{\rm 2P}}}}
 \times 
 e^{i (\lambda_{\ga} \theta_1^{\prime}+
   \lambda\theta_2^{\prime})}
 (-i)^{(\lambda +\lambda_{\gamma})} 
 2\pi 
 J_{(\lambda +\lambda_{\gamma})}(\sqrt{-t_2}b) \; 
 \tilde{J}(\vec{b}, \vec{r}_1,z_1, \vec{r}_2).
 \label{msvfinalampl}
\eeqa
%
%
The resulting differential cross section is:
\beqa
 \frac{d\sigma_{\gamma \, {\rm p}}^{\mathbb{O}}}{dt_2}=
 \frac{1}{16 \pi s_2^2}\frac{1}{2}\sum_{\lambda}
 \sum_{\lambda_{\gamma}} | T_{\lambda,\lambda_\gamma} |^2.
 \label{diffcrossmsv}
\eeqa
%
%
One of the features of the result (\ref{diffcrossmsv}) 
is, that the differential cross
section in the forward direction ($t_2=0$) does not vanish due to the
appearence of the Bessel function $J_0$
in (\ref{msvfinalampl}) for $\lambda+\lambda_{\gamma}=0$. 
The differential cross section (\ref{diffcrossmsv}) for 
photoproduction is displayed in Fig. \ref{msvbild}.
%
%
%
%
%
%
%
%
The integrated cross section is:
\be
\sigma_{\gamma \, {\rm p}}^{\mathbb{O}}
(\gamma {\rm p} \to \pi^0 \{2{\rm P}\}) = 294 \;{\rm nb}.\nn
\ee
%
%
As this photoproduction cross section is constant, i.e. independent of $s_2$,
the EPA conversion to electroproduction can be achieved by simply multiplying
it with a constant $c_{{\rm EPA}} = 0.0136$ corresponding to the $y$ 
integration 
in (3). This gives
\be
\sigma^{\mathbb{O}}
({\rm e p} \to {\rm e} \pi^0 \{{\rm 2P}\}) =  4.01 \;{\rm nb}.\nn
\ee
%
%
The other possibility, namely treating the final state $X$ in
(\ref{profile}) as a free 
quark-diquark pair, is achieved by approximating the final state by a plane 
wave. This is effectively invoking quark-hadron duality, and should be a 
reasonable approximation if we are not interested in the local form of the 
mass spectrum of $X$ but only in the integral over the full mass spectrum. 
In this case the integrated cross section was found to be \cite{donaru}
\be
\sigma_{\gamma {\rm p}}^{\mathbb{O}}
(\gamma {\rm p} \to \pi^0 {\rm X}) = 341 \; {\rm nb}\nn
\ee
%
%
consistent with (14). 

A good check for the use of quark-hadron duality in diffractive
processes is provided by the application of a similar method to the
reaction $\gamma {\rm p} \to \rho {\rm X}$ 
and its comparison with elastic $\rho$
photoproduction. One obtains in the model \cite{doku}:
\be
{\sigma( \gamma {\rm p} \to \rho {\rm p})} = 7.9 \;\mu{\rm b}
\ee
%
%
in good agreement with the pomeron contribution \cite{dl95}
at $W=20$ GeV to the total $\rho$ photoproduction cross section
\cite{ast82}.
For the ratio one obtains\footnote{We thank S. Weinstock for 
communicating this result prior to publication.}
\be
\frac{\sigma( \gamma {\rm p} \to \rho {\rm X})}
{\sigma(\gamma {\rm p} \to \rho {\rm p})}  \approx 1.5.
\ee
%
%
in agreement with the data at HERA energies \cite{weinexp}.

It should be noted that the results for diffractive dissociation depend 
much more on the choice of the wave functions than for elastic processes.
In the latter the overlap becomes essentially the density and is constrained 
by normalization. The photon-pion overlap has been tested to some 
extent by the pion radiative decay, but there is no such test for 
the proton-2P overlap. The odderon contribution is also much more
sensitive to the parameters of the MSV than pomeron exchange.
We have also applied the matrix cumulant expansion technique 
\cite{naber}, and with the approximations as done there we
found results differing from the above ones at most by 50 \%.
Taking all these uncertainties into account we
estimate the uncertainty in the cross section in this 
model calculation to be at least a factor 2 at $W = 20$ GeV.

Finally we return to the question of energy-dependence.
In hadron hadron scattering the increase of the cross  
sections together with the shrinking of the diffraction peak can 
be well reproduced in this model by scaling the hadronic radii by
$(W^2/400 \; {\rm GeV}^2)^{0.08/3}$ \cite{dfk,naber}. Assuming that the same 
radial scaling reproduces the energy dependence of the odderon 
contributions we find that the integrated cross section scales like 
$(W^2/400 \; {\rm GeV}^2)^{0.15}$, 
leading to an enhancement of ca. 1.8 for HERA energies 
as compared to $\sqrt W=20$ GeV.

\section{The electromagnetic cross section}

In this section we consider PS production mediated by photon
rather than odderon exchange (Fig. 1b).
Again the proton, now hit by the photon, is allowed to go 
into some hadronic final state.

The coupling of a pseudoscalar meson PS to two photons is fixed as 
described in \cite{kina}. The Lorenz structure is that of the triangle anomaly
where the strength of the coupling, parametised by a constant $u_{PS}$,
can be extracted from the partial decay width of the PS decaying into
two photons. The scattering amplitude in leading order perturbation theory 
of the electroweak interaction is
\begin{eqnarray}
  &&S_{fi} = i \, (2 \pi)^4 \, \delta^4(p+p_1-p_1'-k-p_X) \, T_{fi},
  \nonumber\\
  &&T_{fi} = -e \, \bar{u}(p_1') \gamma^{\mu} u(p_1) \,
  \frac{1}{t_1} \; u_{PS} \, 
  \epsilon_{\mu \nu \rho \sigma} \, q_1^{\rho} q_2^{\sigma} \, T(t_1,t_2)
  \, \frac{1}{t_2}
  \nonumber\\
  &&\hphantom{T_{fi} = }
  \times \langle X(p_X) | e \, J^{\nu}(0) | P(p) \rangle.
  \label{amplitude}
\end{eqnarray}
%
%
Here $J^{\nu}$ is the hadronic part of the electromagnetic 
current\footnote{Throughout this Section all notations are as in
\cite{nabuch} } and the form factor $T(t_1,t_2)$ is given in terms of 
a vector meson dominance ansatz (cf. \cite{kina}). 

We are not interested in properties due to 
polarisation of the initial and final state particles and so average 
over the initial particle polarisations and sum over the final ones.
In addition we are also not interested in momentum distributions of 
the outgoing hadrons except, clearly, for the PS. For this reason we 
perform the integrations over their momenta. 

Summing over all kinematically accessible states X the differential 
cross section can be written as 
\begin{eqnarray}
  &&d^6 \sigma^{\gamma} = \frac{1}{2\, w(s,m^2,m_e^2)} \, 
  \frac{d^3p_1'}{(2 \pi)^3 \, 2 \, p_1^{' 0}}
  \frac{d^3k}{(2 \pi)^3 \, 2 \, k^{0} } \;\;
  \rho^{\mu \nu} \, P_{\mu \nu},
  \nonumber\\
  &&\rho^{\mu \nu} := \frac{e^2}{t_1^2} \,
  \Big\{ (g^{\mu \nu} - \frac{q_1^{\mu}q_1^{\nu}}{q_1^2}) +
  \frac{ (2 p_1 - q_1)^{\mu}(2 p_1 - q_1)^{\nu}}{q_1^2} \Big\},
  \nonumber\\
  &&P_{\mu \nu} := \frac{e^2}{t_2^2} \,
  u_{PS}^2 \, T^2(t_1,t_2)
  \epsilon_{\mu \omega \alpha \beta}
  q_1^{\alpha} q_2^{\beta} 
  \epsilon_{\nu \rho \gamma \delta}
  q_1^{\gamma} q_2^{\delta}  \; (2 \pi) (2\, m_{{\rm p}})
  \, W^{\rho \omega},
  \nonumber\\
  &&w(x,y,z):= (x-(\sqrt{y}+\sqrt{z})^2)^{\frac{1}{2}}
  (x-(\sqrt{y}-\sqrt{z})^2)^{\frac{1}{2}}.
  \label{diffcross}
\end{eqnarray}
%
%
Here $W_{\rho \omega}$
is the usual hadron tensor as defined for example in chapter
18 of \cite{nabuch} involving  the two invariant functions
$W_1$ and $W_2$.
If the sum over all kinematically accessible states is
restricted to some subset invariant 
under Lorentz and parity transformations 
the two invariant functions $W_{1,2}$ which appear in
$W_{\rho \omega}$ change,
whereas the tensor structures, multiplied by $W_{1,2}$, do not.
The photoproduction cross section 
of (3) for photon exchange  $\sigma_{\gamma {\rm p}}(s_2)$ is
given by
\begin{eqnarray}
  &&\sigma_{\gamma {\rm p}}^{\gamma}(s_2) =
  \int \frac{d^3k}{2 k^0 (2 \pi )^3} \,
  \frac{1}{2 \, w(s_2,m^2,t_1)} \, 
  \frac{1}{2} \, ( -g^{\mu \nu}) \,
  \nonumber\\
  && \hphantom{d\sigma_T^{\ast}=}
  (u_{PS}^2 \, T^2 \, e^2 \, 
  \epsilon_{\mu \omega \alpha \beta}
  q_1^{\alpha} q_2^{\beta} 
  \epsilon_{\nu \rho \gamma \delta}
  q_1^{\gamma} q_2^{\delta}) \, \frac{1}{t_2^2}
  \nonumber\\[0.05cm]
  &&\hphantom{d\sigma_T^{\ast}=}
  (2 \, \pi) \,(2 \,  m_{{\rm p}}) \, 
  W^{\rho \omega}.
  \label{sigtotallg}
\end{eqnarray}
%
%
We now evaluate (\ref{sigtotallg})
using the data on total $\gamma^{\ast} {\rm p}$
cross section  collected in \cite{brasse} where
the $\gamma^{\ast} {\rm p}$ c.m. energy is below 
2.0 GeV and $0 \le |t_2| \le 6.0$ GeV$^2$.
As a check on the procedure we compare the result for the $\Delta(1232)$
resonance region with an explicit calculation of $\Delta$ electroproduction.

\subsection{The resonance region}

\begin{figure}[htb]
  \unitlength1.0cm
  \begin{center}
    \begin{picture}(15.,8.8)
      \put(3.0,1.0){
        \epsfysize=9.0cm

        \epsffile{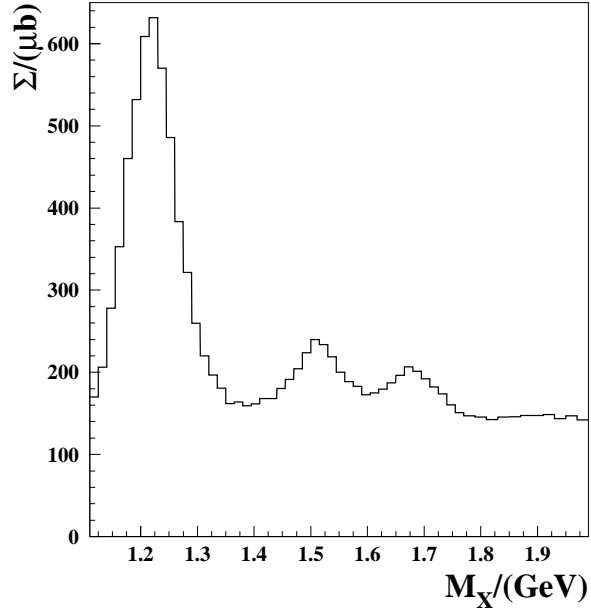}}
    \end{picture}
  \end{center}
  \vspace*{-1.5cm}
  \caption{The $\gamma^{\ast} {\rm p}$ absorption cross section 
 $\Sigma$ for $q_2^2=-0.1$ GeV$^2$ as a histogram using the 
 parametisation (\ref{fit})}
  \label{histobild}
\end{figure}

The total $\gamma^{\ast} {\rm p}$
absorption cross section $\Sigma = \sigma_T + \epsilon\sigma_L$ 
in the region of the main nucleon resonances has been determined in inelastic 
electron proton scattering experiments.  With the kinematical
definitions of \cite{hand} the relation of $\sigma_T$ and $\sigma_L$
to $W_{1,2}$ is:
\begin{eqnarray}
  && W_1 (M_X^2,q_2^2) = \frac{1}{4 \pi^2 \alpha} \,
  \frac{M_X^2-m_{{\rm p}}^2}{2 \, m_{{\rm p}}} \,
    \sigma_T (M_X^2,q_2^2),
    \nonumber\\
    && W_2 (M_X^2,q_2^2) = \frac{1}{4 \pi^2 \alpha} \,
\frac{M_X^2-m_{{\rm p}}^2}{2 \, m_{{\rm p}}} \,
  (1-\nu^2/q_2^2)^{-1} \,
  \big[ \sigma_T(M_X^2,q_2^2) + \sigma_L(M_X^2,q_2^2) \big],
  \nonumber\\
  && \nu = \frac{1}{2 \, m_{{\rm p}}} \, (M_X^2-m_{{\rm p}}^2 -q_2^2).
  \label{kindef}
\end{eqnarray}
%
%
Here $\alpha$ is the fine structure constant.
For very small photon virtualities 
there are photoproduction data in addition. In \cite{brasse} all these data 
have been combined to extract the following parametrisation of $\Sigma$ 
as a function of the photon virtuality $q_2^2$ and the invariant 
mass squared of the final state $M_X^2$ 
\begin{eqnarray}
  &&\ln\Big( \frac{\Sigma}{G_D^2} \Big) = a(M_X^2)+
  b(M_X^2) \ln \Big( \frac{ |\vec{q}_2|}{|\vec{q}_2|_0} \Big) +
  c(M_X^2) | \ln \Big(\frac{ |\vec{q_2}|}{|\vec{q}_2|_0} \Big) |^{d(M_X^2)},
  \nonumber\\
  &&G_D(q_2^2) =  \frac{1}{(1 - q_2^2/0.71 {\rm GeV^2})^2 }.
  \label{fit}
\end{eqnarray}
%
%
%
%
$G_D$ is the dipole form factor, $|\vec{q}_2|$ ($|\vec{q}_2|_0$) is 
the absolute value of the photon momentum
in the proton rest frame for photon virtuality $q^2_2$ ($q^2_2=0$). The fit is 
restricted to the range $1.11 \le M_X \le 1.99$ GeV. The functions $a,b,c$ 
are given in \cite{brasse} only at discrete points\footnote{In the range $1.11
\le M_X \le 1.755$ in steps of $\Delta M_X$=0.015 GeV and in the range $1.755 
\le M_X \le 1.990$ in steps of $\Delta M_X$=0.02 GeV.}. The function $d$ is 
taken to be a constant, $d = 3.0$. To convert the fit to a continuous function
in $M_X^2$ and $q_2^2$ we represent $\Sigma$ as a histogram.  
Fig. \ref{histobild} shows 
as an example $\Sigma$ for a fixed photon virtuality of $q_2^2=-0.1$ GeV$^2$.

%
It was shown in \cite{brasse} that the total virtual photon-proton cross 
section for longitudinal photons $\sigma_L$ is small in the resonance region,
so we neglect it in our discussion. Setting thus $\sigma_T=\Sigma,\,
\sigma_L=0$ in (\ref{kindef})
we get an analytic expression for the hadron tensor $W_{\mu \nu}$
and are in a position to calculate the inelastic PS production cross 
section (\ref{sigtotallg}), summing over all hadron final states $|X \rangle $ 
with invariant mass $1.11\le M_X \le 1.99$ GeV. 

As a specific check on our procedure we shall want to focus on a particular
mass range, namely that of the $\Delta(1232)$ resonance. This can be achieved 
as follows. We multiply $\sigma_{\gamma {\rm p}}(s_2)$ in (\ref{sigtotallg})
by

\begin{eqnarray}
  1 = \int  d \, M_X^2 \; \delta(p_X^2 - M_X^2)
  \label{trick}
\end{eqnarray}
%
%

and define 
\begin{eqnarray}
  &&\tilde{\sigma}_{\gamma {\rm p}}^{\gamma} (s_2) :=
  \int \, \frac{d^3k}{2 k^0 (2 \pi )^3} \,
  \frac{1}{2 \, w(s_2,t_1=0,m_{{\rm p}}^2)} \, 
  \frac{1}{2} \, ( -g^{\mu \nu}) \,
  \nonumber\\
  && \hphantom{d\sigma(s_2,t_2)=}
  ( 4 \pi \alpha \, u_{PS}^2 \, T^2  \, 
  \epsilon_{\mu \omega \alpha \beta}
  q_1^{\alpha} q_2^{\beta} 
  \epsilon_{\nu \rho \gamma \delta}
  q_1^{\gamma} q_2^{\delta} )\, \frac{1}{t_2^2}
  \nonumber\\[0.05cm]
  &&\hphantom{d\sigma(s_2,t_2)=}
  (2 \, \pi) \, (2 \, m_{{\rm p}}) \, W^{\rho \omega}(p_X^2,t_2) \,
  m_{{\rm p}}^2 \, \delta(p_X^2-m_X^2) 
  \label{newsigma}
\end{eqnarray}
%
%
%
Contracting the Lorentz indices in (\ref{newsigma}), inserting the explicit
form of the hadron tensor and expressing the invariant functions through 
(\ref{kindef})
we get finally for the cross section of inelastic 
PS production 
\begin{eqnarray}
  &&\sigma^{\gamma} = 
  \int_{M_{min}}^{M_{max}} \, d M_X \; \frac{2 \, M_X}{m^2} \,
  \int_{y_{min}}^{y_{max}} \, \frac{d y}{y} \; n(y) \;
  \int_{t_{2_{min}}}^{t_{2_{max}}} \, d t_2 \; 
  \frac{d \tilde{\sigma}_{\gamma {\rm p}}^{\gamma}}{d t_2} (s_2),
  \nonumber\\[0.1cm]
  &&
  \frac{d \tilde{\sigma}_{\gamma {\rm p}}^{\gamma}}{d t_2}=
  \frac{m^2 T^2 u^2 \; \Sigma(M_X^2,t_2)}
  {64 \pi^2  ( s_2 - m_{{\rm p}}^2 )^2 t_2^2}\,
  \frac{M_X^2 - m_{{\rm p}}^2}
  {m_{{\rm p}}^4 + (M_X^2-t_2)^2 - 
  2 m_{{\rm p}}^2 ( M_X^2 + t_2)}
  \nonumber\\ 
  &&\hphantom{\frac{d \tilde{\sigma}_T}{d t_2}}
  \Big\{  m_{PS}^4( m_{{\rm p}}^2 - M_X^2 )^2 
  - 2 t_2 m_{PS}^2 \big(
  M_X^2 ( m_{PS}^2 + M_X^2 - s_2) +
  m_{{\rm p}}^2 ( s_2 - M_X^2 ) \big) +
  \nonumber\\
  &&\hphantom{m_{PS}^4( \{ \frac{d \tilde{\sigma}_T}{d t_2}}
  t_2^2 \big( m_{{\rm p}}^4 + m_{PS}^4 + M_X^4 +
  2 m_{PS}^2 ( 2 M_X^2 - s_2 ) + 2 m_{{\rm p}}^2 ( m_{PS}^2 - s_2) -
  \nonumber\\
  &&\hphantom{ff m_{PS}^4(  \{ t_2^2 \big( \frac{d \tilde{\sigma}_T}{d t_2}}
  2 M_X^2 s_2 + 2 s_2^2 \big) +
  2 t_2^3 \big( s_2 - m_{{\rm p}}^2 - m_{PS}^2 - M_X^2 \big)
  + t_2^4 \; \; \Big\}
  \label{epax}
\end{eqnarray}
%
%
%
The integration limits with respect to $t_2$ are just the
kinematical limits of the two by two process 
$\gamma + {\rm p} \rightarrow {\rm PS + X}$
as given,
for example, in \cite{particle}.

In the second row of Tab. \ref{tot} we list our results 
for the total cross section of
inelastic PS electroproduction in the photoproduction region 
(\ref{cuts}), calculated from (\ref{epax}).
\begin{table}[hbt]
\begin{center}
$
\begin{array}{|l||c|c|}
  \hline
  & \sigma^{\gamma}({\rm e \, p}\rightarrow {\rm e\, PS\, X}) 
  & \sigma^{\gamma}({\rm e \, p}\rightarrow {\rm e\, PS\, p})\\
  \hline
  \pi^0 & 2.0 \; {\rm pb} & 78.1 \; {\rm pb} \\
  \hline
  \eta  & 1.9 \; {\rm pb} & 56.4 \; {\rm pb} \\
  \hline
  \eta^{\prime} & 3.1 \; {\rm pb} & 83.6 \; {\rm pb} \\
  \hline
  \eta_c & 0.3 \; {\rm pb} & 3.83 \; {\rm pb}\\ 
  \hline
\end{array}
$
\caption{Total cross section for for inelastic and elastic PS
  electroproduction by photon exchange.}
\label{tot}
\end{center}
\end{table}
These values are compared in the third row of Tab. \ref{tot}
to the results of  
elastic PS production calculated in terms of the
EPA in \cite{kina}.
As we can read off from
Tab. \ref{tot} the contributions from the inelastic PS production
integrated cross sections 
are very small compared to the elastic ones. However 
in experimental  analyses one typically has to make additional cuts
on the data and, of course, this will change the ratio of inelastic to
elastic contributions. We explore this first for the $t_2$-distribution.

\begin{figure}[htb]
  \unitlength1.0cm
  \begin{center}
    \begin{picture}(15.,6.8)
      \put(3.5,0.0){
        \epsfysize=8.0cm

        \epsffile{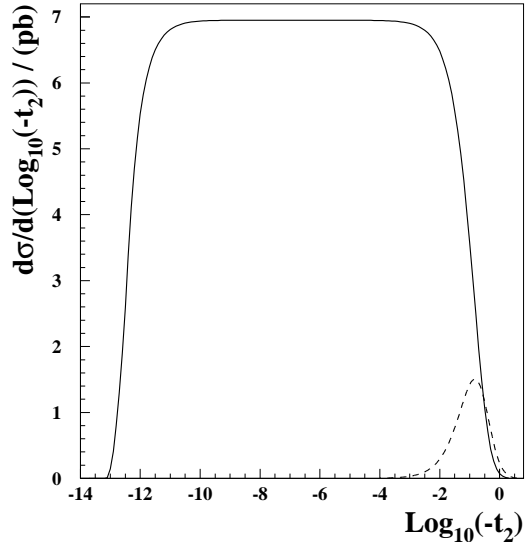}}
    \end{picture}
  \end{center}
  \vspace*{-0.4cm}
  \caption{The ${\rm Log}_{10}(-t_2)$ distributions for elastic
    $\pi^0$
    production (solid line) 
    taken from [22]
    compared to case of inelastic
    production (dashed line), integrated over the 
    whole resonance region ($1.11 \le M_X \le 1.99$ GeV). }
\label{t2bild}
\end{figure}

\begin{figure}[htb]
  \unitlength1.0cm
  \begin{center}
    \begin{picture}(15.,7.8)
      \put(-0.5,0.0){
        \epsfysize=8.50cm
        \epsffile{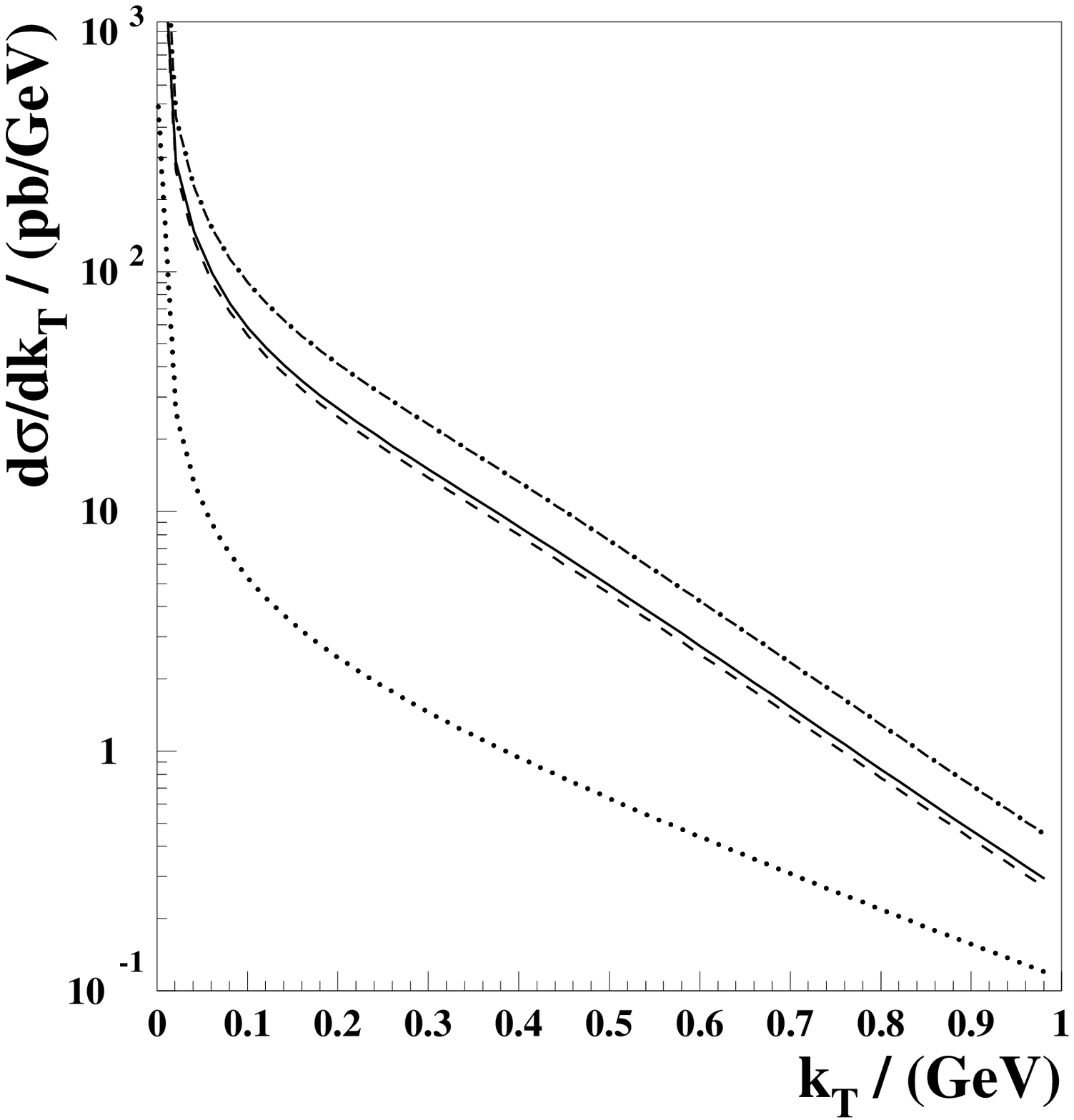}}

      \put(7.5,0.0){
        \epsfysize=8.50cm
        \epsffile{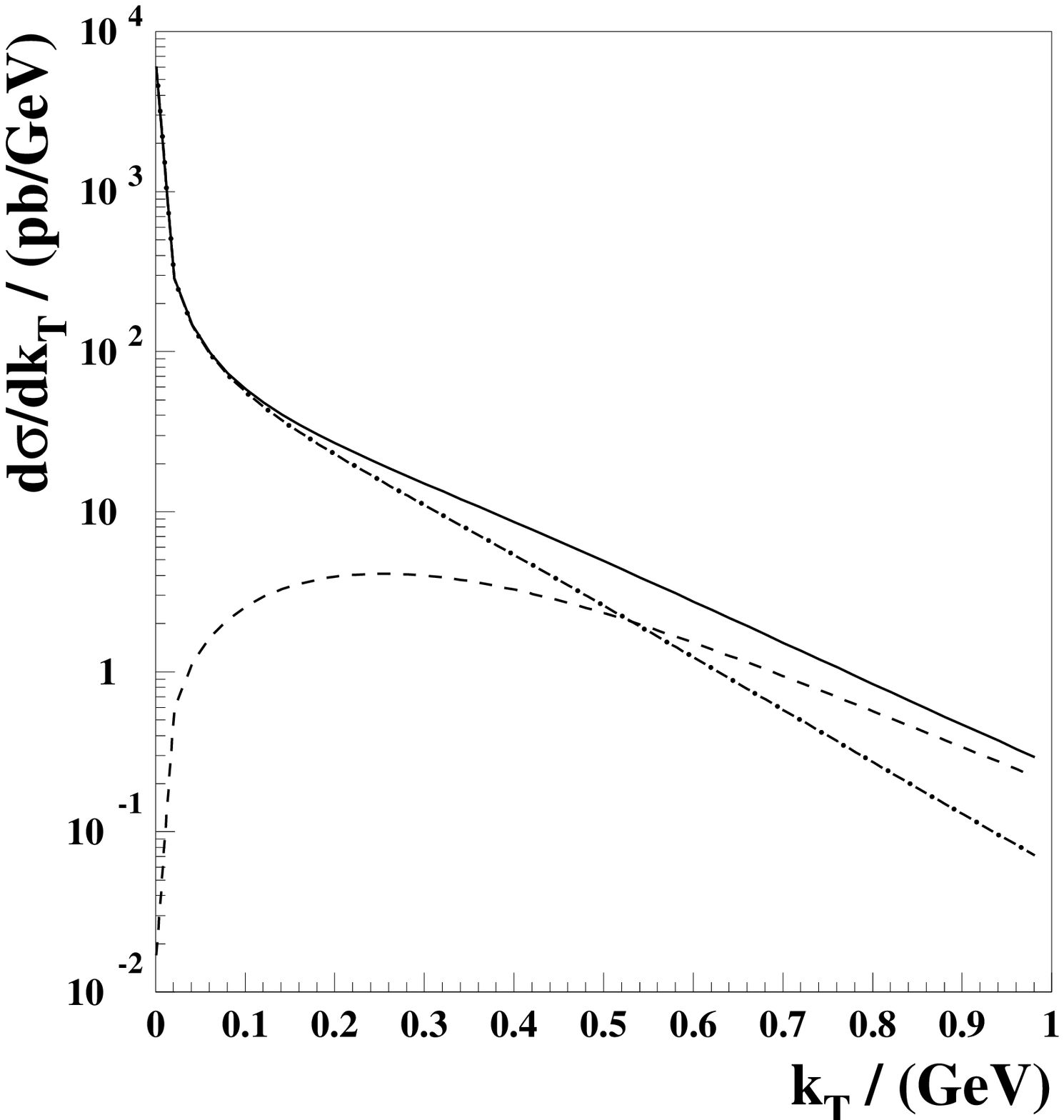}}

    \end{picture}
  \end{center}
  \vspace*{-0.0cm}
  \caption{(a) The complete electromagnetic result for the
    $k_T$ distributions of  $\pi^0$
    (solid line), $\eta$ (dashed line),  $\eta'$
    (dashed dotted line) and $\eta_c$ (dotted line) production.
    (b) The elastic (dashed dotted line) and the inelastic
    contribution
    (dashed line) together with the full electromagnetic
    result for pion production}
\label{ktbild}
\end{figure}

To compare with the results of
\cite{kina} it is convenient to plot the differential cross section with 
respect to the logarithm of $-t_2$ rather than 
$d\sigma^{\gamma}/dt_2$. Our results 
for $\pi^0$ production are shown in Fig. \ref{t2bild}, together with the 
results of \cite{kina} for the elastic case. 
The cross sections for the other pseudoscalar mesons
in essence scale as the coupling constants $u_{PS}^2$. 
It is immediately clear 
that the large difference in integrated cross section 
between the elastic and inelastic case is coming from the 
region of very small $|t_2|$. In the limit $t_2 \rightarrow 0$ 
$d\sigma^{\gamma}/dt_2$ 
goes to a constant for the break-up calculation and
simultaneously the available phase space region in  
$t_2$ becomes smaller and 
smaller. This results in the decrease of the logarithmic distributions 
in Fig. \ref{t2bild} for $|t_2|\le 10^{-2}$. 
In the case of the elastic production $d\sigma^{\gamma}/dt_2$ increases 
as $1/|t_2|$ as $t_2 \rightarrow 0$. 
The corresponding logarithmic distribution  
then shows a plateau which is cut off by the upper phase space limit of 
$t_2$ \cite{kina}. 
%
%
%
%
%
%
%
Another and experimentally preferred variable  to measure is
$|\vec{k}_T|\equiv k_T$,
the transverse momentum of the pseudoscalar mesons, defined relative to the 
beam axis. At HERA $k_T$ distributions can be measured for values of $k_T$ 
greater than $O$(0.1 GeV). 
The photoproduction cuts of (\ref{cuts}) restrict the 
transverse 
momentum of the incident photon to be smaller than $O$(0.1 GeV), so in 
our case there is 
practically no distinction between the beam axis and the photon
axis. Then within the accuracy of the
EPA calculations we have $t_2 = -k_T^2$.
The complete electromagnetic result 
$d\sigma^{\gamma}/dk_T$ 
for all four pseudoscalars is shown in Fig. \ref{ktbild}.
We see that applying a cut $k_T > 0.1$ GeV
the break-up cross section represents a much larger
fraction of the total
then for the integrated cross sections (Table 1).
%
%
%
%
%
%

\subsection{The $\Delta$(1232) resonance}

In this section we consider the hadron final state 
$|{\rm X} \rangle$ to be the resonance $\Delta$(1232), and evaluate the
cross section in two ways. Firstly we use the procedure of Section 3.1
but restrict ourselves to the $\Delta$ region 
which we define as $1.11 \le M_X \le 1.40$ GeV (see Fig. \ref{histobild}).
Secondly the $\Delta$ is treated in the isobar model \cite{isobar}
as a stable particle (zero width approximation) of spin 3/2. 

A spin 3/2 particle can be described \cite {rarita} by a vector-spinor 
$R^{\mu}$, which is contained in the direct product of the vector and the 
Dirac representation of the Lorenz group:

\begin{eqnarray}
  R^{\mu} \; \epsilon \; (\frac{1}{2},\frac{1}{2})
  \otimes \big[ (\frac{1}{2},0) \oplus (0,\frac{1}{2}) \big].
  \nonumber
  \label{rari}
\end{eqnarray}
To project out the spin 3/2 part of this reducible representation
corresponding to the $\Delta$-particle
one has to require $\gamma_{\mu}R^{\mu}$=0 and 
$p_{\Delta}^{\mu}R_{\mu}$=0 \cite{rarita,weinberg}
and that every component of $R^{\mu}$ fulfills 
the Dirac equation $(p_{\Delta}.\gamma-m_{\Delta})R_{\mu}$=0. Let
$R_{\mu}(p_{\Delta},i)$  (i=1,...,4)  be a set of normalised basis 
vector-spinors:
\begin{eqnarray}
  \bar{R}_{\mu}(p_{\Delta},i) R^{\mu}(p_{\Delta},j) = \delta_{ij} \; \; \; \; 
  i,j=1,...,4.
  \label{norm}
\end{eqnarray}
%
%
%
To calculate the partial width or the total and differential cross sections 
we only need the polarisation sum $S_{\mu \nu}$. Using the above conditions  
we get for the polarisation sum
($v_{\Delta}=p_{\Delta}/M_{\Delta}$)

\begin{eqnarray}
  S^{\mu \nu}(p_{\Delta}) &=& \sum_i R^{\mu}(p_{\Delta},i) 
\bar{R}^{\nu}(p_{\Delta},i)
  \nonumber\\
  &=& \frac{1}{3} \Big( \frac{\gamma.v_{\Delta} + 1}{2} \Big)
  \big(3 g^{\mu \nu} - 2 v_{\Delta}^{\mu} v_{\Delta}^{\nu} - 
  \gamma^{\mu} \gamma^{\nu} \big) 
  \Big( \frac{\gamma.v_{\Delta} + 1}{2} \Big).
  \label{spinsum}
\end{eqnarray}
%
%

\begin{figure}[htb]
  \unitlength1.0cm
  \begin{center}
    \begin{picture}(15.,6.8)

      \put(-0.7,0.0){
        \epsfysize=8.0cm
        \epsffile{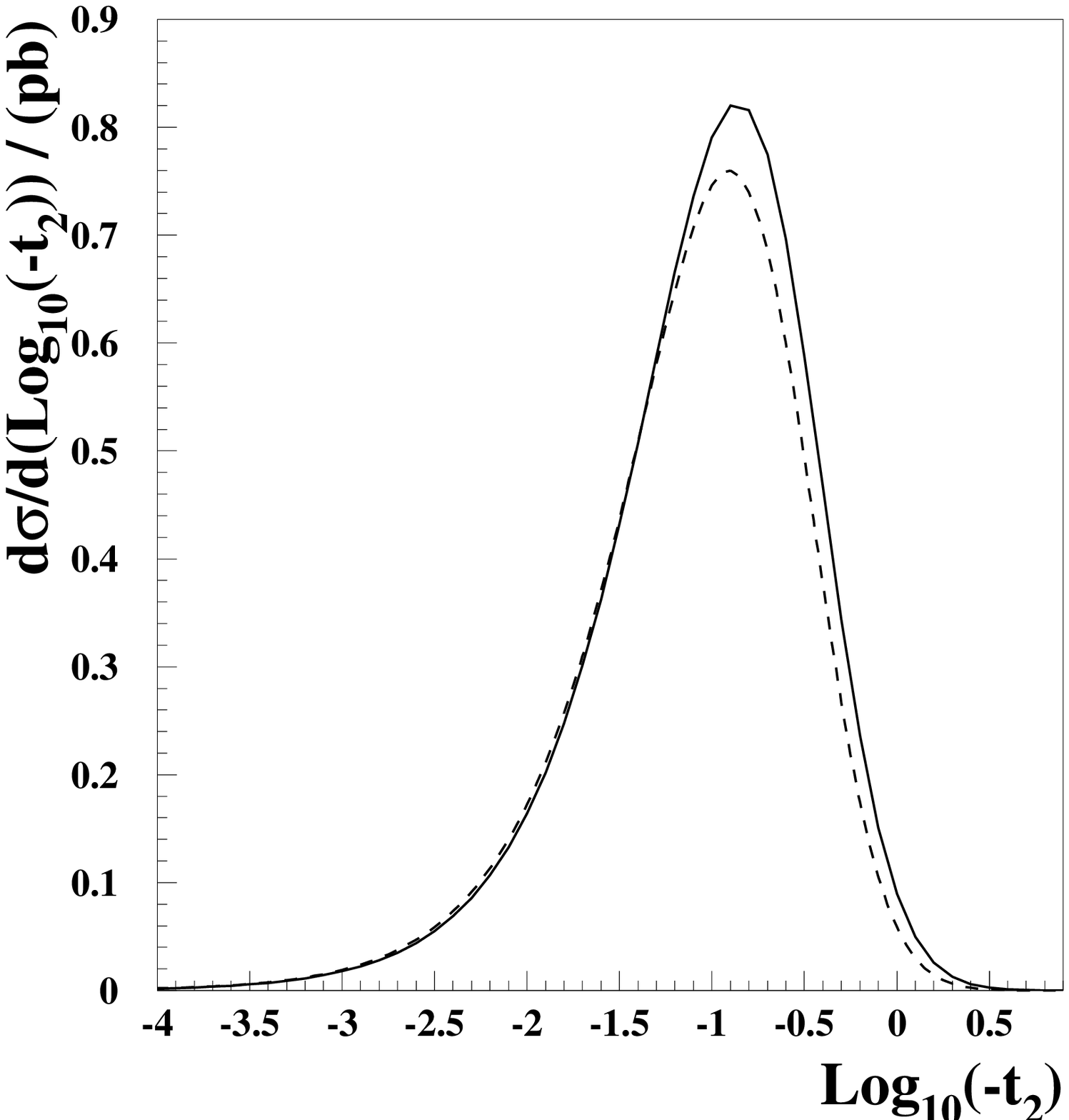}}

      \put(8,0.0){
        \epsfysize=8.0cm
        \epsffile{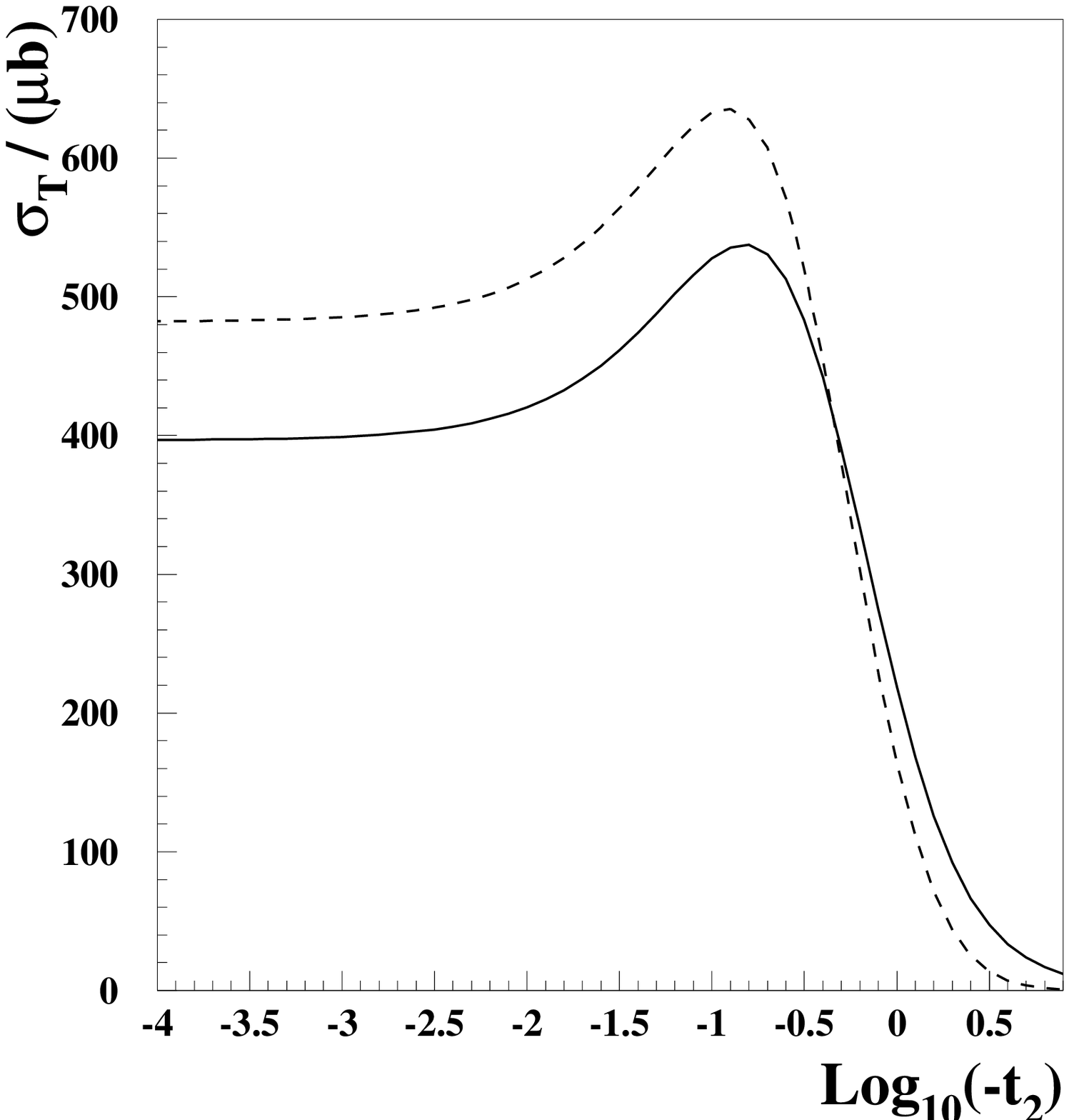}}

    \end{picture}
  \end{center}
  \vspace*{-0.3cm}
  \caption{(a) ${\rm Log}_{10}(-t_2)$ distributions for the delta
    resonance region, 
    calculated from the 
    $\gamma^{\ast} {\rm p}$
    cross section data as explained in 
    Sect. 3.1, integrated over the
    invariant mass range $1.11 \le M_X \le 1.4$ GeV (dashed line),
    and from the isobar model (solid line). 
    (b) $\sigma_T=\Sigma$ at $M_X=1.215$ GeV 
    from (\ref{fit}) as a function of $t_2$
    (dashed lines)
    compared with $\sigma_T$ calculated from $W_1^{\Delta}$.}
  \label{compare}
\end{figure}

The coupling of the $\Delta$ resonance to the proton and the 
photon is parametrised through
\begin{eqnarray}
  \frac{2 e f_{\gamma R}}{M_{\Delta}} \epsilon_{\mu \nu \rho \sigma}
  p_{\Delta}^{\rho} p^{\sigma} \, G_D(q_2^2),
  \label{deltac}
\end{eqnarray}
%
%
%
which corresponds to the covariant formulation of the 
isobar model \cite{isobar}. Here $G_D$ is defined in (23) and
$f_{\gamma R}$ is the coupling constant. From (\ref{deltac})
together with (\ref{spinsum}) we can calculate the $\Delta$
decay into a real photon and a proton. The result is
\begin{eqnarray}
  \Gamma (\Delta^+ \rightarrow {\rm p} \gamma) =
  \frac{e^2 f_{\gamma R}^2}{3 \, \pi} \;
  \frac{(M_{\Delta}^2-m_{{\rm p}}^2)^3}{8\, M_{\Delta}^5}.
  \label{ddecay}
\end{eqnarray}
%
%
%
This partial decay width is known experimentally \cite{dalitz,berends},
$\Gamma (\Delta^+$ $\rightarrow$ $p \gamma)$= $0.65 \pm 0.02$ MeV.
Taking the central value
from which it follows that $f_{\gamma R}$=2.4. Now everything is fixed
and we 
can easily derive the hadron invariant functions 
$W_1^{\Delta},W_2^{\Delta}$: 

\begin{eqnarray}
  W_1^{\Delta} &=& \frac{f_{\gamma R}^2 G_D^2}{3 M_{\Delta}^4 m_{{\rm p}}^2}
  ((p q_2)^2 - p^2 q_2^2 ) \, f(pq_2,t_2) \,
  \delta ( (p+q_2)^2 - M_{\Delta}^2), 
  \nonumber\\
  W_2^{\Delta} &=& \frac{f_{\gamma R}^2 G_D^2}{3 M_{\Delta}^4 m_{{\rm p}}^2}
  (-p^2 q_2^2) \, f(pq_2,t_2) \,
  \delta ( (p+q_2)^2 - M_{\Delta}^2), 
  \nonumber\\
  f(p q_2, t_2) &=& \{ \,
  (m_{{\rm p}} + 2 M_{\Delta}) m_{{\rm p}}^2 + 
2(m_{{\rm p}} + M_{\Delta}) p q_2 
  + m_{{\rm p}} ( M_{\Delta}^2 + t_2) \; \}.
  \label{wd}
\end{eqnarray}
%
%
%
Having derived the hadron tensor it is then simply a task of 
contracting indices to get the cross section of photon proton
scattering into a PS and the $\Delta^+$ and, by applying the EPA, the 
total electroproduction cross section. 
\begin{eqnarray}
  &&\sigma^{\gamma} = \int_{y_{min}}^{y_{max}} \, 
  \frac{dy}{y} \, n(y) 
  \int_{t_{2_{min}}}^{t_{2_{max}}} \, 
  dt_2 \, \frac{d \sigma_{\gamma {\rm p}}^{\gamma}}{dt_2} (s_2,t_2),
  \nonumber\\
  &&\frac{d \sigma_{\gamma {\rm p}}^{\gamma} }{d t_2}=
  \frac{e^2 f_{\gamma R}^2 G_D^2 u_{PS}^2 T^2 }
  {384 m_{{\rm p}} M_{\Delta}^3 \pi 
    (s_2 - m_{{\rm p}}^2 )^2 t_2^2 }
  \big( (m_{{\rm p}} + M_{\Delta})^2 - t_2 \big)
  \nonumber\\
  &&\hphantom{\frac{d \tilde{\sigma}_T}{d t_2}}
  \Big\{  m_{PS}^4( m_{{\rm p}}^2 - M_{\Delta}^2 )^2 
  - 2 t_2 m_{PS}^2 \big(
  M_{\Delta}^2 ( m_{PS}^2 + M_{\Delta}^2 - s_2) +
  m_{{\rm p}}^2 ( s_2 - M_{\Delta}^2 ) \big) +
  \nonumber\\
  &&\hphantom{m_{PS}^4( \{ \frac{d \tilde{\sigma}_T}{d t_2}}
  t_2^2 \big( m_{{\rm p}}^4 + m_{PS}^4 + M_{\Delta}^4 +
  2 m_{PS}^2 ( 2 M_{\Delta}^2 - s_2 ) + 2 m_{{\rm p}}^2 ( m_{PS}^2 - s_2) -
  \nonumber\\
  &&\hphantom{ff m_{PS}^4(  \{ t_2^2 \big( \frac{d \tilde{\sigma}_T}{d t_2}}
  2 M_{\Delta}^2 s_2 + 2 s_2^2 \big) +
  2 t_2^3 \big( s_2 - m_{{\rm p}}^2 - m_{PS}^2 - M_{\Delta}^2 \big)
  + t_2^4 \; \; \Big\}
  \label{dsigdelta}
\end{eqnarray}
%
%
%
This calculation is compared with the result 
for $d\sigma/d {\rm Log}_{10}(-t_2)$
of Sect. 3.1,
restricted to the $\Delta$ region in Fig. \ref{compare}a.
 %
%
%
%
%
The comparison is quite satisfactory. The deviations can be
understood qualitatively by plotting  $\sigma_T=\Sigma$ 
from (\ref{fit}) at
$M_X=1.215$ where the curve 
in Fig. \ref{histobild} has its maximum,
as a function of $t_2$ together with $\sigma_T$
calculated from $W_1^{\Delta}$ in (\ref{wd}) using the definition
(\ref{kindef}) and replacing
the $\delta$-function in (\ref{wd}) with $1/(\pi M_{\Delta}\Gamma)$
where $\Gamma$ is the total width of the $\Delta$, 
$\Gamma \approx 120$ MeV \cite{particle} (Fig. \ref{compare}). 
For $|t_2| \le 0.1 \; {\rm GeV}^2$ the fit (\ref{fit}) to the measured 
cross section $\Sigma$ is somewhat larger due to a 
significant S-wave component which is absent in the calculation of the
$\Delta$ alone but the shapes are nearly the same. For  
$|t_2|\ge 0.1 \; {\rm GeV}^2$ the model calculation of $\sigma_T$
decreases too slowly compared to experiment. In essence this translates
directly into the shapes of Fig. \ref{compare}a where for 
$|t_2| \le 0.1 \; {\rm GeV}^2$ the 
differential cross section with respect to the logarithm of
the exact result is 
a little bit larger than the one of the model calculation whereas for 
$|t_2| \ge 0.1 \; {\rm GeV}^2$  the model calculation
overshoots the exact one.

\section{Conclusions}

We have calculated pseudoscalar electroproduction by odderon exchange
with nucleon dissociation in a specific model which has proven successful
in processes dominated by pomeron exchange. The principal conclusion
is that the cross section is significantly larger for nucleon breakup
than when the nucleon 
remains intact. 
%
%
\begin{figure}[htb]
  \unitlength1.0cm
  \begin{center}
    \begin{picture}(15.,7.4)
      \put(2.5,0.0){
        \epsfysize=9.0cm

        \epsffile{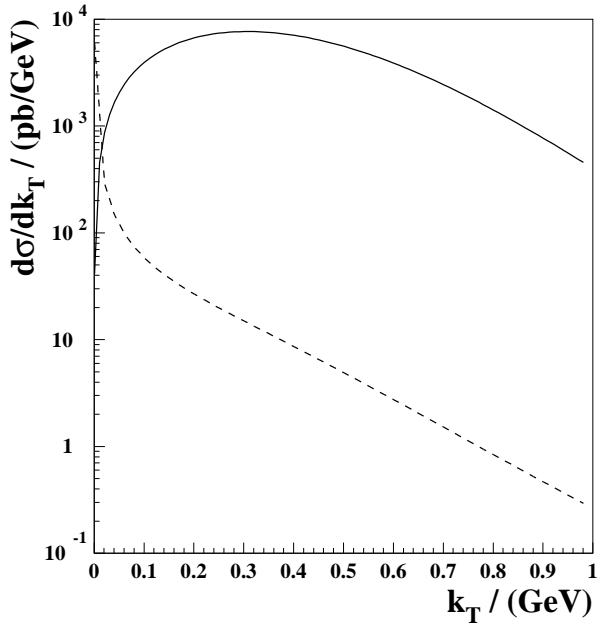}}
    \end{picture}
  \end{center}
  \vspace*{-0.0cm}

  \caption{The $k_T$ distribution in pion production
for odderon exchange (solid line)
calculated from the amplitude 
(\ref{msvfinalampl}) of Sect. 2  compared to the
complete electromagnetic result (dashed line). Interference 
contributions are not taken into account.}
\label{k_Tmsv}
\end{figure}
%
%
%
We have also presented the results of an ``exact'' calculation of the same
process by photon exchange. When combined with the previous  
calculation for the elastic case
\cite{kina} we have a precise prediction for the purely
electromagnetic cross section.

In Fig. \ref{k_Tmsv} we show for $\pi^0$ production the
distribution of the pion's $k_T$ summed over the elastic and 
inelastic channels  for the electromagnetic and the odderon 
exchange. Interference terms are not taken into account here.
Even taking the most pessimistic view of the uncertainties 
in the model the process should be observable at HERA. 
This would establish the soft odderon as an exchange-object
in high energy scattering on equal footing with the soft pomeron.
On the other hand
failure to observe any significant
deviation from the electromagnetic result
would be clear evidence for the 
complete failure of our model. 
One possible conclusion would then be that the soft
odderon really does not exist and
that our understanding of diffractive processes is much
less than is believed.

The competing hadronic process, Reggeised $\omega$ exchange, can be estimated 
from what is known about $\pi^0$ photoproduction at low energies, 
$\sqrt s \le 5.5$ GeV
and the falloff with energy given 
by Regge theory:
$d \sigma / d t_2|_{\omega-exch} \sim s^{2 \,
  \alpha_{\omega}(t_2)-2}$
with $\alpha_{\omega}(0) \approx$ 0.18 to 0.5.
The result is very much smaller than the calculated 
odderon contribution and is not a serious background to it. However it 
could conceivably be sufficiently large to interfere with the photon exchange
amplitude and so simulate a very weakly-coupled odderon. 

\section{Acknowledgements}

The authors would like to thank P. V. Landshoff, K. Meier, T. Berndt,
S. Tapprogge, A. Likhoded, W. Buchm\"uller, E. Meggiolaro, M. Strikman,
B. Nicolescu, A. Hebecker, H. J. Pirner, G. Kulzinger and S. Weinstock 
for useful discussions and correspondence.

\end{document}